# Photonic Dipole Contours of Ferrofluid Hele-Shaw Cell

Michael Snyder, Jonathan Frederick

*Murray State University*

*Department of Engineering and Physics*

*Murray State University, Murray, KY 42071*

*sirzerp@gmail.com*

## Abstract

This investigation describes and demonstrates a novel technique for the visualization of magnetic fields. Two ferrofluid Hele-Shaw cells have been constructed to facilitate the imaging of magnetic field lines. We deduce that magnetically induced photonic band gap arrays similar to electrostatic liquid crystal operation are responsible for the photographed images and seek to mathematically prove the images are of dipole nature.

A simple way of explaining this work is to think of the old magnetic iron filling experiments; but now each iron filling is a molecule floating in a liquid. Each molecule has the freedom to act as an independent lens that can be aligned by an external magnetic field. Because each lens directs light, the external field can be analyzed by following the light paths captured in the photographs.

## I. Introduction

Understanding magnetic fields is important to facilitate magnetic applications in industry, commerce, and space exploration. Electromagnets can move heavy loads of metal. Magnetic materials attached to credit cards allow for fast and accurate business transactions. Magnetic fields are not visible, and therefore often hard to understand or characterize.

The basic idea of our experiment is to analyze and understand the lines that we have seen and photographed in a very thin layer of ferrofluid placed within an external magnetic field. Ferrofluid is a fluid containing dispersed nanoscale magnetic particles. Each particle is a single magnetic domain which is colloidally stabilized to prevent agglomeration. In other words, each ferrofluid particle has a magnetic moment and a small electrostatic charge to stop them from clustering together. Ferrofluid is commonly used in loud speakers to increase the speaker coil's heat dissipation and power ratings, and also used in magnetic bearing seals in industry.

A Hele-Shaw cell consists of two flat plates that are parallel to each other and separated by a small distance, and at least one of the plates is transparent. They are mostly used in chemistry and fluid dynamics to study fluid viscosity and density gradients. We believe the first use of a Ferrofluid Hele-Shaw cell was by R. E. Rosensweig in the book Ferrohydrodynamics [1], to investigate the microscopic physical proprieties of ferrofluids.

We seek to analyze the paths that the light follows when injected orthogonally into a ferrofluid Hele-Shaw cell (also referred to as "the lens" or "the cell" in this paper). Each cell is made of two circular, optically flat, windows of glass sandwiched together with a very thin layer of ferrofluid in between. Light is injected radially into the edges at regular intervals, as seen in Figure 1. We define the flat plane of the glass as the xy plane; the light was injected into the outside edges of the glass, and the pictures were taken by a camera on the z axis.

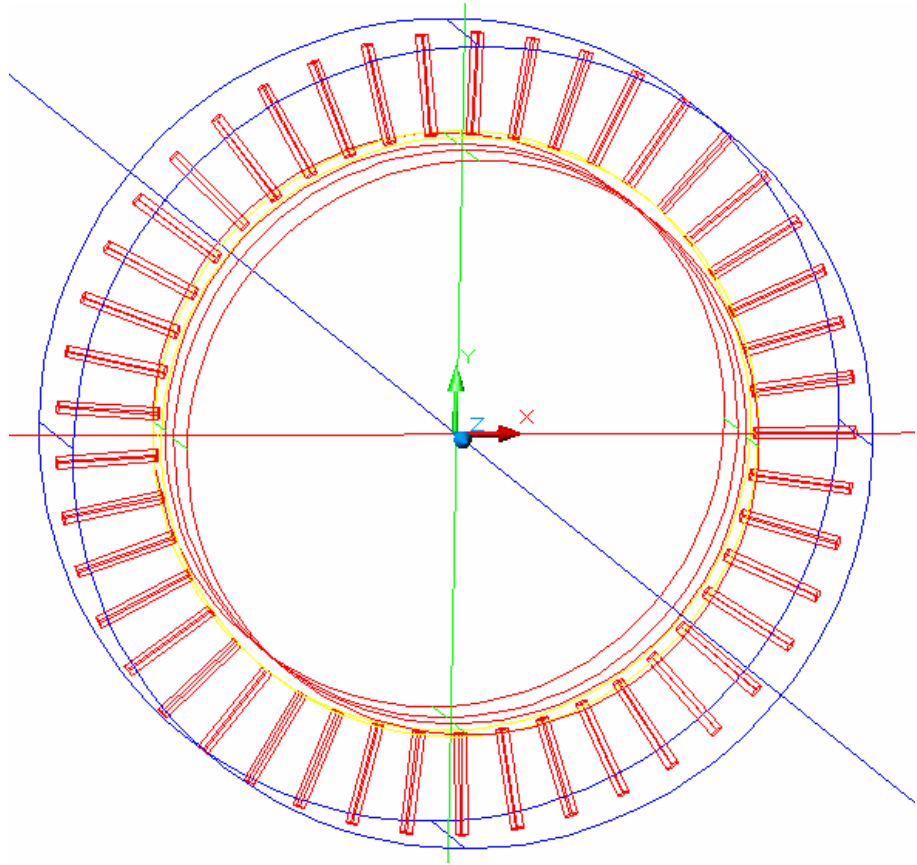

Figure 1: Diagram of lens mount. The radial holes are for light injection.

Two versions of the apparatus were utilized. One was made up of two 150mm diameter BK7 glass windows with a parallelism of 1 arc minute. The second was made up of two 114mm diameter windows that had a hole of 38mm removed from the centers. Both were filled with commercial ferrofluid EFH1 which uses light mineral oil as a medium. We estimate the fluid layer is roughly a micron in thickness because the windows have a rated optical flatness of ¼ wavelength at 650nm, which is liberally rounded up to be 500nm per side when the windows are butted and glued together. Optical glue was used on the edges of the glass windows and capillary action used to draw the ferrofluid inside each cell.

The magnets used for the photographs were 25.4mm diameter spherical neodymium magnets mounted on wood dowels and rotated by computer controlled stepper motors. Each neodymium magnet was factory rated to be equal or greater than a tesla in strength and because we were not solving for magnetic permeability of the ferrofluid medium, the strength measurement was not critical for our experiment. We found that spherical magnets gave the smoothest field images. The experimental setup is shown in Figure 5 and Figure 6.

## II. Background information

All measurements were kept in pixels. Our first goal was to identify the dipole nature contained within our photographs. To define a dipole one needs only the location of one of the poles and the distance vector to the other pole.

The ferrofluid Hele-Shaw cells were made by Timm Vanderelli from Ligonier, Pennsylvania.  Mounts and lighting systems were created to investigate his claim of an optical presentation of magnetic flux [2]. Both fiber optics and light emitting diodes have been used to inject photons into the lenses. Figure 2 is a microscopic image [3] of chains of ferrofluid particles flowing to a magnet chip located on a microscope slide

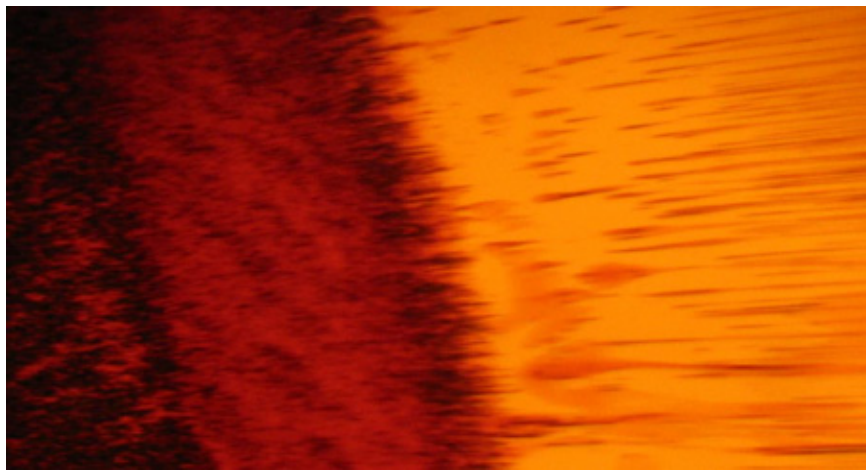

Figure 2: Image of ferrofluid particles in the presence of a static magnetic field.

The illumination leads to two different color response modes of the lenses. Starting at red wavelengths and working to green wavelengths, photons pass through the lens giving detailed dipole contour lines of the externally imposed magnetic field, as seen in Figure 3.

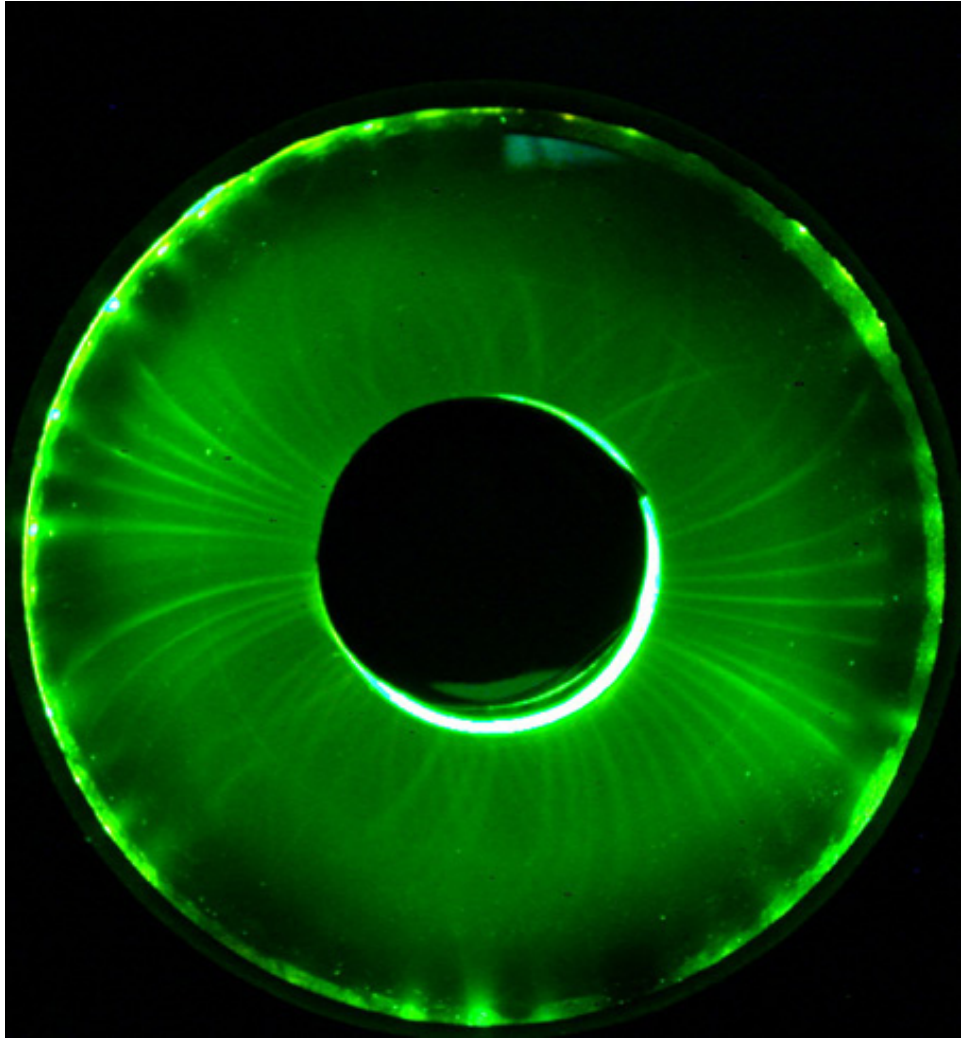

Figure 3: Image taken with green light and vertical dipole alignment.

Blue wavelengths provide significantly different pictures. The blue light images tend to form an evenly highlighted background, and imposing a magnetic field seems to restrict the amount of light oriented toward the viewer, particularly near the pole locations, leaving a uniform blue field with black void features. The pole areas seen in Figure 9 and Figure 10 are similar as the ones first seen with the blue wavelength responses.

Lasers were found to create images but tended to give irreproducible results. Most likely this is due to the limitations of our equipment. Using a green laser pointer, we were able to find general features of our pictures without having the LEDs of the apparatus turned on. In other words, if one used a laser pointer along the z axis and pointed at the cell with an external field present; one could find individual features such as a pole or single line which could be illuminated across the cell, when part of the feature was highlighted by the laser. Surprisingly, if a polarization filter was placed between the laser and the cell, one could tune the filter until the feature disappeared while the laser was still aimed at the same location. We found that directed but diffused LED light works best for our apparatus.

## III. Mathematical Background

Because our experiment had no free current other than the ferrofluid particles slowly moving in the oil based medium, we modeled the curl of the external B field as equaling zero ($\nabla \times B = 0$) which allowed us to model the images with ($B = -\mu \nabla V_M$) a magnetic scalar potential. It was observed that the light seemed to follow isopotential lines of the scalar field. To create a mathematical form of isopotential lines, or lines of constant value, it is useful to model our magnets as two magnetic charges producing a magnetic dipole.

$$V_m = \frac{\mu M_c}{4\pi} \left[ \frac{1}{r_n} - \frac{1}{r_s} \right] \tag{1}$$

In equation (1) $V_m$ is the magnetic potential, $\mu$ is the permeability of the ferrofluid medium, $M_c$ is a fictional magnetic charge, and $r_n$, $r_s$ are the vector locations of the north and south poles. Because it is only the geometry of the images that we are seeking to analyze, we will consider the $\frac{\mu M_c}{4\pi}$ constant as equaling one.

$$\frac{1}{k} = \frac{1}{\sqrt{(x-x_n)^2 + (y-y_n)^2 + (z-z_n)^2}} - \frac{1}{\sqrt{(x-x_s)^2 + (y-y_s)^2 + (z-z_s)^2}} \quad (2)$$

Equation (2) should be recognized as a well known field dipole equation in the Cartesian coordinate system. In this case $V_m$ has been replaced as $\frac{1}{k}$, a given constant. The reciprocal form of $k$ was chosen for reasons of symmetry. Each value of k produces a unique set of curved contour lines. What might not be recognized is that equation (2) is equivalent to a high degree polynomial without a general solution.

$$\text{Let } x_n, y_n, z_n = 0, \frac{d}{2}, 0 \quad \& \quad \text{Let } x_s, y_s, z_s = 0, -\frac{d}{2}, 0 \quad (3)$$

Show in Equation (3), let the north pole be $\frac{d}{2}$ distance above the origin and let the south pole be $-\frac{d}{2}$ distance below the origin on the y axis. Equation (4) shows the substitutions. Equation (5) has other substitutions that will help solve for a specific solution. Equations (6-10) are algebraic manipulations leading to the polynomial form.

$$\frac{1}{k} = \frac{1}{\sqrt{x^2 + (y-\frac{d}{2})^2 + z^2}} - \frac{1}{\sqrt{x^2 + (y+\frac{d}{2})^2 + z^2}} \quad (4)$$

$$\text{Let } p = x^2 + (y-\frac{d}{2})^2 + z^2 \quad \& \quad \text{Let q} = x^2 + (y+\frac{d}{2})^2 + z^2 \quad (5)$$

$$\left[\frac{1}{k}+\frac{1}{\sqrt{q}}\right]^2 = \left[\frac{1}{\sqrt{p}}\right]^2 \tag{6}$$

$$\frac{1}{k^2}+\frac{2}{k\sqrt{q}}+\frac{1}{q}=\frac{1}{p} \tag{7}$$

$$\left[\frac{2}{k\sqrt{q}}\right]^2 = \left[\frac{1}{p}-\frac{1}{q}-\frac{1}{k^2}\right]^2 \tag{8}$$

$$0 = \frac{1}{k^4}-\frac{2}{k^2 p}-\frac{2}{k^2 q}+\frac{1}{p^2}-\frac{2}{pq}+\frac{1}{q^2} \tag{9}$$

$$0 = p^2 q^2 - 2k^2 pq(p+q) + k^4 (p-q)^2 \tag{10}$$

Finally, equation (11) is a solvable specific form with of our field dipole equation. We have verified that the Maple computer algebra system can solve equation (11) for single variables.

$$0 = (x^2 + (y-\frac{d}{2})^2 + z^2)^2(x^2 + (y+\frac{d}{2})^2 + z^2)^2 - 2k^2(x^2 + (y-\frac{d}{2})^2 + z^2)(x^2 + (y+\frac{d}{2})^2 + z^2)$$
$$*(2x^2 + (y-\frac{d}{2})^2 + (y+\frac{d}{2})^2 + 2z^2) + k^4((y-\frac{d}{2})^2 - (y+\frac{d}{2})^2)^2 \tag{11}$$

## IV. Data Acquisition

The still photographs were taken with a Nikon 995 digital camera and the movies with a CoVi CVQ-2110 security camera. The CoVi provided an excellent low light NTSC output that was captured onto our lab computer. The Nikon 995 is one of the better low cost cameras for scientific purposes because its exposure and focus settings are lockable during operation providing for a consistent series of photographs. We used a Nikon Wide Angle WC-E63 lens in order to gather as much light as possible. Our best still images came from an exposure setting of 8 seconds and a 4.6 F-stop.

The still photographs were calibrated by using standard astronomical techniques of taking five dark frames before and after every run. The dark frame average was then subtracted from each photograph as it was processed within Matlab.

Our processors of choice were Matlab for still images and the program VirtualDub by Avery Lee for the video editing [4]. The standard processing was to break up the photographs and videos into RGB layers and process each layer separately. Contrast and lighting settings were chosen to enhance the dipole lines contained within the images. At no time was information added to photographs; but video filters were used to remove unwanted information. For example, in VirtualDub, the black & white, smoothing, and level filters were used in the videos [5].

In Matlab the still photographs were broken up into RGB arrays, with dark frame noise subtracted. Then contrast was set by multiplying the image arrays by a scalar value. A radial light intensity, distance dependence was found within the photos.

The centers of the photographs were brighter than the light sources at the edges. This made digitizing the images challenging because near the center there was one signal to noise ratio, but the levels just a few hundred pixels away were different, meaning any digitizing operation in Matlab would fail because the photos did not have consistent signal and noise levels.

This challenge was overcome by multiplying the value of each pixel by 1.02 times its pixel distance away from the center. This equalization process allowed us to create binary masks from the photographs.

## V. Data Analysis

Observation of the resulting images suggested that the lines were most likely isopotential lines of a scalar magnetic potential. To confirm this we chose a parameter variation method to find the dipole locations. Our method was to examine the first 24 photographs of a particular run and guess the possible locations of the poles. From this set of initial conditions, a calculation of the associated magnetic field for that pole location could be produced. A computer algorithm was written to vary the locations of the initial guess in an iterative process until a best fit was obtained in comparing the simulation to the observed set of field lines.

To generate the idealized dipole field plots, an array was created and filled with inverse values radiating away from the center. To make a dipole plot, a smaller section of the pole array was taken and then subtracted from an offset section of the same array. The resulting array is a dipole field array. To make it easier to work with we deleted the near field values and used the absolute values of the array. The deleted near field values produced the black circles in Figure 4.

The computer then tagged regions of equal field magnitude values. For example, in Matlab one can issue a single command that finds all the .002 values in the complete array and changes them into $10^{99}$ values. Then one can digitize all values greater than $10^{98}$ to ones and all else to zeros. This produces a very quick and accurate isopotential line image.

In order to identify those isopotential lines of interest, it was decided to seed the ideal dipole image with equal field magnitude locations from the photographs. This is equivalent to an array indexing operation in a sense, to find a white spot on the original, and then look at the same location on the candidate image array and find the value underneath it. We then tagged all those values in the idealized image equal to the seed-referenced value as seen in Figure 4. Since both the original and idealized images are made up of curves, the chances of tagging a wrong line and finding an image comparison fit is very small. If one tags the incorrect line, it simply curves away from the area of interest.

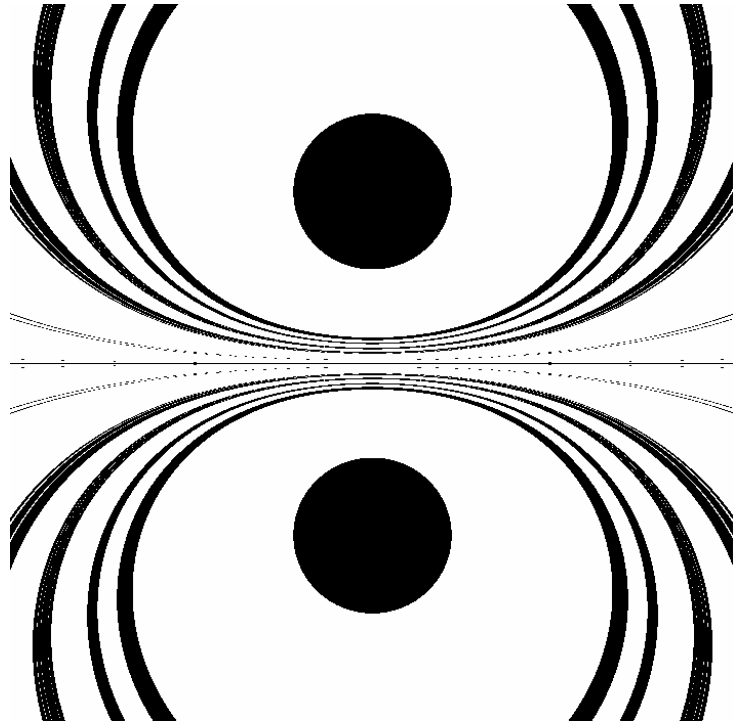

Figure 4: This is an example of the isopotential models.

Only when both the computer generated pole parameters and the correct isopotential lines are tagged does the computer return a best fit. The best fit is the fit that most closely resembles the experimental image, with the most tagged points. See Figures S1-S2 and F1-F2 for examples of the fits.

## VI. Results & Discussion

After using Matlab to find the best **d** for 24 experimental images we found that:

**d**=325±9 pixels (95%)

This value is larger than the approximate photographed diameter of 215 pixels for our spherical magnets. This places the fictional magnetic charges in free space about 55 pixels above and below the physical magnet. In other words where the traditional B field lines would diverge from the poles; our fictional magnetic pole charges exist in our photographs. Table 1 shows our measured **d** values. The **x** and **y** values are the best pole location fit, that was determined by Matlab. Variable **d** is just the pixel distance from $(x_1, y_1)$ to $(x_2, y_2)$.

One resulting question from our experiments is how does one take a picture of isopotential contour lines? As we proved in equation (11); such curved lines are not trivial and could not be accidentally photographed.

The images only behave as light following a dipole surface in a limited area of the lens. In the areas where the north and south poles of the magnet are located, crossing of the light paths is seen. This is contradictory to a simple dipole model. We suggest that the crossing lines are related to left-handed and right-handed polarizations of the light inside the cell.

Figure 7 through Figure 12 are photographs from other datasets. In Figure 7 the blue LED's of the 150mm cell has been replaced with 15 sets of sequenced red, yellow, and green LED's. A tesla rated 25.4mm cube magnet has been placed at the bottom edge of the cell with a pole facing the camera. Notice how the color bands warp around the magnet, and sometimes combine into new colors. A circle band of white light is seen adjacent to the physical magnet. The isopotential lines around a pole should be perfect circles, but at the edge of the cell the circles are clearly distorted.

There are other pole pictures showing a trend of the circle shape distortion decreasing as the magnet approaches the center of the cell. In other words, there seems to be a minimum turn radius before the injected light in the cell can align itself with the potential of the external magnetic field.

In Figure 9, two cube magnets are photographed. The first magnet on the left is aligned north-south and the second on the right is aligned south-north. This places different poles at the top and bottom of the image. Twisting helical lines are seen in the center between the left and right magnets. Away from the center, parallel contour lines are seen between all four poles.

In Figure 10, the same two cube magnets are photographed. In this case both are aligned with the same poles at the top and bottom. Horizontal mostly parallel lines are seen between the left and right sides of the cell. The enclosed dark areas representing the poles have coupled together and have apparently induced a third set of poles between the two magnets.

After working with the cells for two years and processing thousands of photographs and hours of videos [5], the lines seen in the cells seem to follow some basic properties. Different poles (ns or sn) have perpendicular lines between the poles representing magnetic potential. Same type poles (nn or ss) have helical lines connecting the poles together. The enclosed dark areas representing the poles as seen in the blue led photographs and the three color led photographs, are loosely coupled to the physical magnets that produce them and can be distorted by their magnetic environment. Furthermore, same type pole areas ostensibly can induce weaker pole areas between themselves.

The visual effects seen in the cells appear not to distinguish between physical poles of the same magnet nor physical poles of different magnets nor induced poles between magnets. The profile view of a single magnet has contour lines between its different poles, and two different poles directly facing the camera of two different magnets also produce a similar set of contour lines.

In Figure 11, the photograph is of a zoomed feature also seen in the center of Figure 9. We find this feature unique and highly structured, and therefore we have given this structure the name of a 'Point of Inversion' and have defined it as the geometric shape of taking a spider web between your hands and flipping one side of the web 180 degrees. Mathematically this feature appears in our photographs when ($r_{n1} - r_{s2} = r_{s1} - r_{n2}$) meaning that it is exactly in the center of the pair of dipoles with one dipole inverted in relation to the other dipole. If one studies this location in Figure 11, one should notice that we have pairs of different color lines originating at this center point and furthermore the line pairs cross over each other onward to different sides of a pole. One obvious conjecture is that each pair is made up of left-handed and right-handed polarizations of light, and each pair is crossing over itself in the presentence of a magnetic field and that the separation observed is Stern–Gerlach in nature.

In Figure 12, there is a 50.8mm ring magnet with the north pole facing the camera. The repeating light pattern produces a nice spirograph. Upon further study, one should notice that each repeating shape is a 'D' shape with the straight part of the 'D' near the center of the magnet and the curved part of the 'D' close to the inside diameter of the magnet. We have found using a green laser pointer, that you could shoot the laser light near the inside diameter of the ring magnet and only get a single 'D' shaped light pattern. The apparent conjecture is much the same as in Figure 11, that left-handed and right-handed polarizations of light originate close to the inside diameter of the magnet and travel in different directions to the center of the magnet forming a spirograph pattern.

In Figure 8, the photograph is showing a different optical effect. A uniform white light source has been placed 30 degrees off the xz axis. The main light source is off screen, and behind and to the right of the cell. Three tesla rated magnets with the north poles facing the camera are photographed. The off axis lighting shows a helix structure between the poles. We believe that narrowest point(s) of the helix between the north poles are related to Points of Inversion.

Table 1: Results of parameter variation in Matlab.

| Image | $x_1$ | $y_1$ | $x_2$ | $y_2$ | d |
|---|---|---|---|---|---|
| 1 | 547 | 706 | 869 | 701 | 322.0 |
| 2 | 547 | 709 | 871 | 709 | 324.0 |
| 3 | 547 | 704 | 871 | 709 | 324.0 |
| 4 | 547 | 704 | 870 | 709 | 323.0 |
| 5 | 541 | 706 | 870 | 702 | 329.0 |
| 6 | 542 | 704 | 870 | 704 | 328.0 |
| 7 | 546 | 705 | 865 | 701 | 319.0 |
| 8 | 547 | 707 | 865 | 703 | 318.0 |
| 9 | 542 | 702 | 869 | 707 | 327.0 |
| 10 | 540 | 702 | 871 | 708 | 331.1 |
| 11 | 547 | 704 | 871 | 708 | 324.0 |
| 12 | 547 | 705 | 871 | 702 | 324.0 |
| 13 | 547 | 709 | 867 | 702 | 320.1 |
| 14 | 546 | 709 | 871 | 709 | 325.0 |
| 15 | 539 | 707 | 871 | 707 | 332.0 |
| 16 | 547 | 701 | 864 | 709 | 317.1 |
| 17 | 547 | 708 | 871 | 706 | 324.0 |
| 18 | 547 | 709 | 871 | 709 | 324.0 |
| 19 | 539 | 706 | 871 | 706 | 332.0 |
| 20 | 546 | 708 | 871 | 701 | 325.1 |
| 21 | 539 | 707 | 871 | 701 | 332.1 |
| 22 | 539 | 703 | 863 | 702 | 324.0 |
| 23 | 539 | 703 | 863 | 702 | 324.0 |
| 24 | 539 | 707 | 868 | 706 | 329.0 |

## VII. Theories of Operation

We have searched for publications involving the optical properties of ferrofluids have found a good number of related papers that seem to be applicable to our experiment:

1) Novel Convective Instabilities in a Magnetic Fluid by Weili Luo and Tengda Du.

2) Improved formulas for magneto-optical effects in ferrouids by Mircea Rasa.

3) Optical properties of magnetic and non-magnetic composites of ferrofluids by Rajesh Patela, R.V. Upadhyaya, and R.V. Mehta.

4) Photonic-crystal resonant effect using self-assembly ordered structures in magnetic fluid films under external magnetic fields by S.Y. Yang, H.E. Hornga, Y.T. Shiaoa, Chin-Yih Hongb,and H.C. Yangc.

5) Light-Bending Nanoparticles by Nikolay A. Mirin, and Naomi J. Halas.

We have listed the papers to show that we have read the published research and were unable to understand how to apply their principles and results to our macro-scale experiment. Instead we wish to share our thoughts about how the ferrofluid cells could be explained in general terms.

The ferrofluid cells are simple instruments in the sense they only have three moving parts. The first moving part is the ferrofluid particles, free to rotate in three dimensions and to move within the limited volume of the cell. The second moving part is the photons that travel through out the cell. The third moving part is the virtual photons that make up the magnetic fields that interact with the ferrofluid particles.

The first hypothesis is that ferrofluid medium is the primary reason for the light paths seen in cells. The act of applying an external magnetic field supplies energy to the medium and allows it to form a complex liquid crystal lattice. Photons play a passive role and merely follow the created channels within the self-assembled structures.

The second hypothesis is the ferrofluid medium is a primary reason for the patterns seen in the cells but also the photons are active participants within the cell. Photons of left-handed and right-handed polarizations can pick different paths through the magnetically induced liquid crystal.

The third hypothesis is that medium does not form a liquid crystal but each molecule in the liquid has a magnetic moment and the external alignment of the molecules can influence the direction of the photons within the medium without being part of a lattice. Each molecule has the freedom to act as an independent lens and molecule to molecule relaying of light can literally cause photon paths within the cells to be circles.

The fourth hypothesis is that the virtual photons are the primary reason for the photon paths seen in the cell. The large amount of virtual photons flowing through the cell aligns the ferrofluid particles for the smallest amount of their cross-sectional area. The instrument acts as a photon bubble chamber in the sense that the medium is saturated with photons. When a channel is formed by the virtual photons then the opportunistic photons simply travel in the created paths. Thus the observed Stern–Gerlach separations are properties of the virtual photons.

## VIII. Conclusions

Ferrofluid Hele-Shaw cells are useful as a magnetic field visualization instrument, unlike any other known to the authors. It is quite conceivable that other utilizations will be found with further research. Many questions have been raised during our current research including the magnetic field strength required for the phenomenon to occur and the actual mechanism by which it occurs.

We have shown that the non-crossing light paths seen in the ferrofluid cells can be approximated by the scalar isopotential lines of the external magnetic field. We have characterized the crossing lines observed in the cells as Stern–Gerlach separations.

The mechanism we propose is one of magnetically induced photonic band gap arrays similar to electrostatic liquid crystal operation. The main idea being that the rod like Ferrofluid molecules dynamically line up head to tail like reflective compass needles and form three dimensional waveguides that direct the radially injected light along the surfaces of a magnetic dipole.

## References & Links

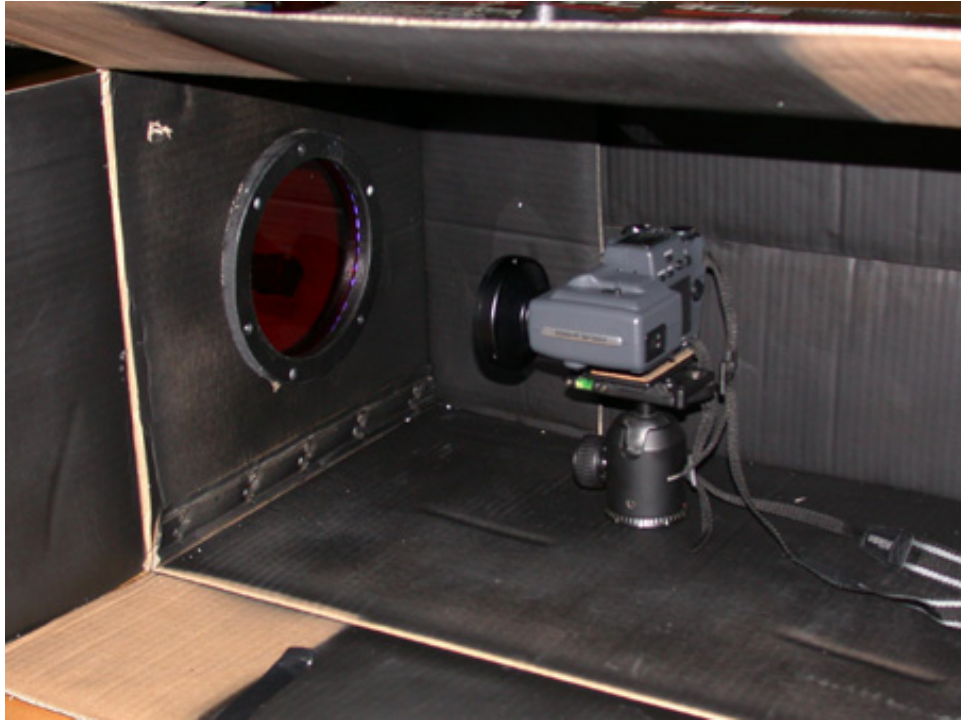

Figure 5: The experimental setup showing the camera and ferrofluid cell on the left face of the box. During operation, the box is closed, and the room lights usually turned off.

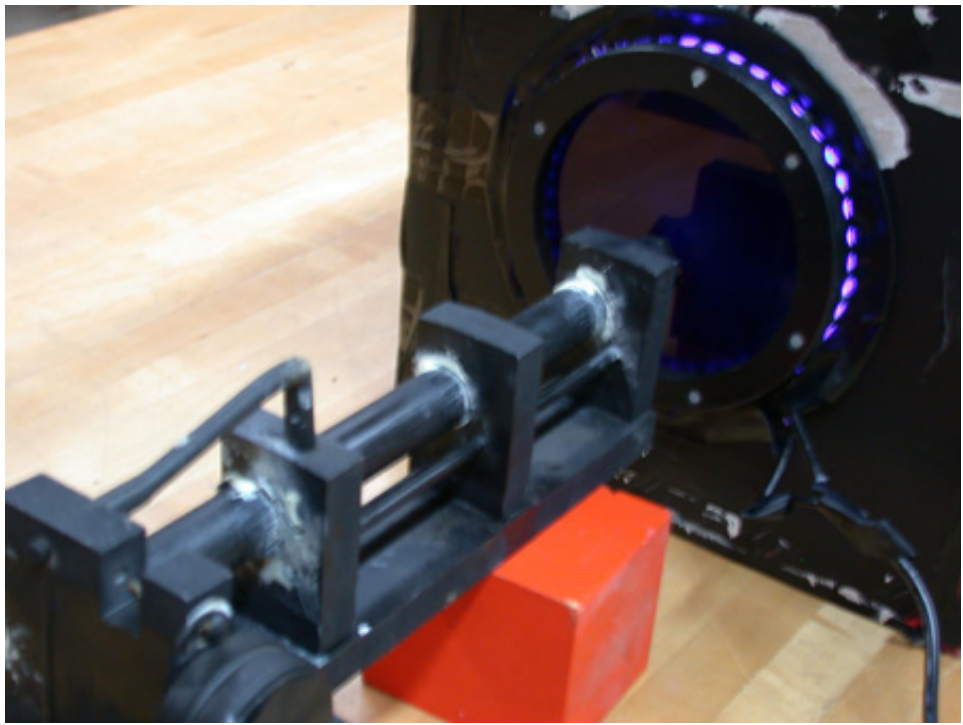

Figure 6: The apparatus showing the magnet location and use of stepper motor mounted on a PVC frame. Blue LEDs were used in this case

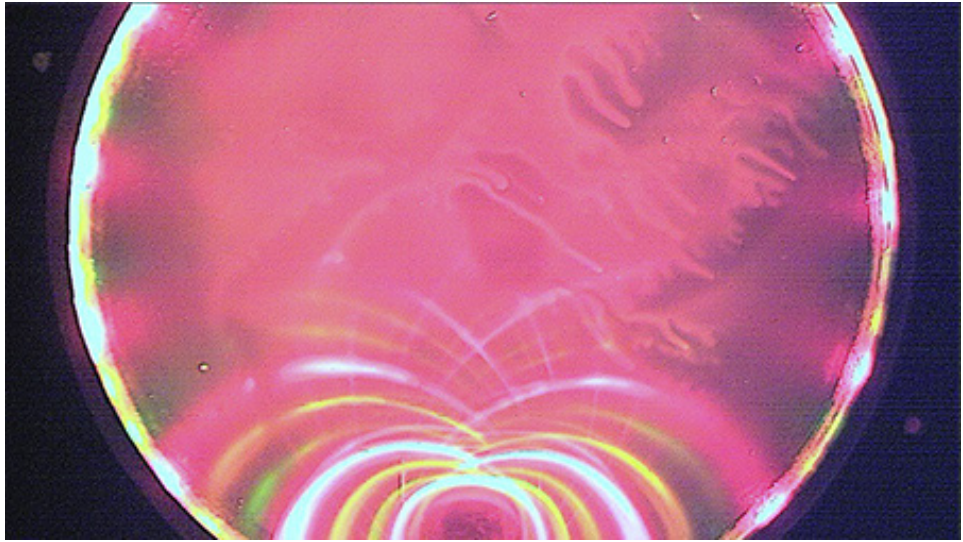

Figure 7: Photograph of a 25.4mm cube magnet at the edge of the lens. Pole facing the camera with different color LEDs around the perimeter of the lens.

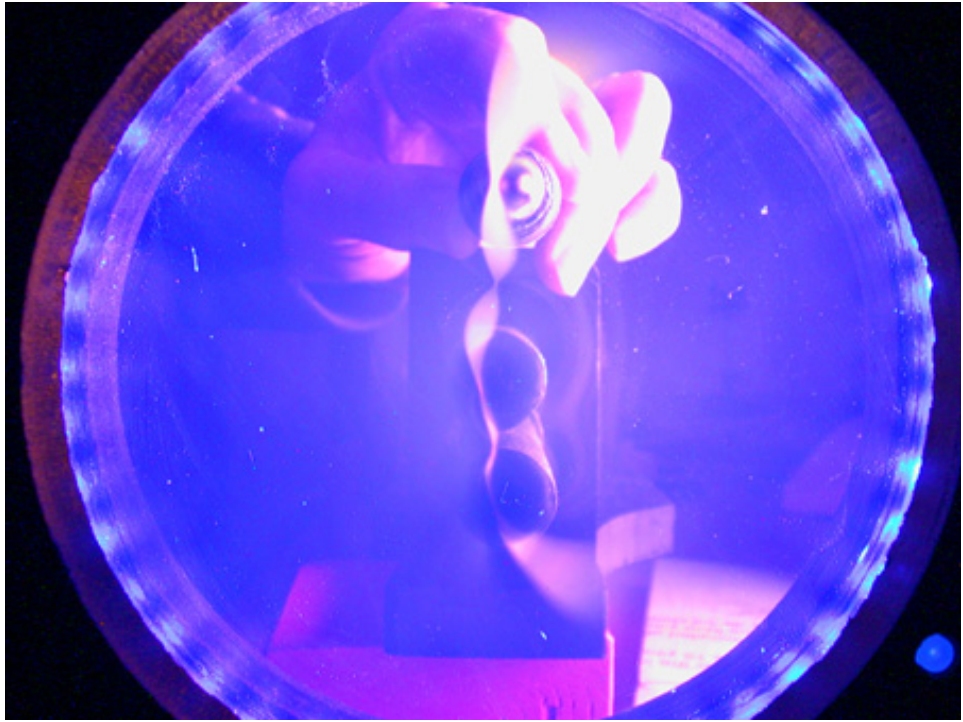

Figure 8: Photograph of an off axis lighting effect with white light source to the right (not visible) and blue LEDs turned on. Three of the same poles facing camera.

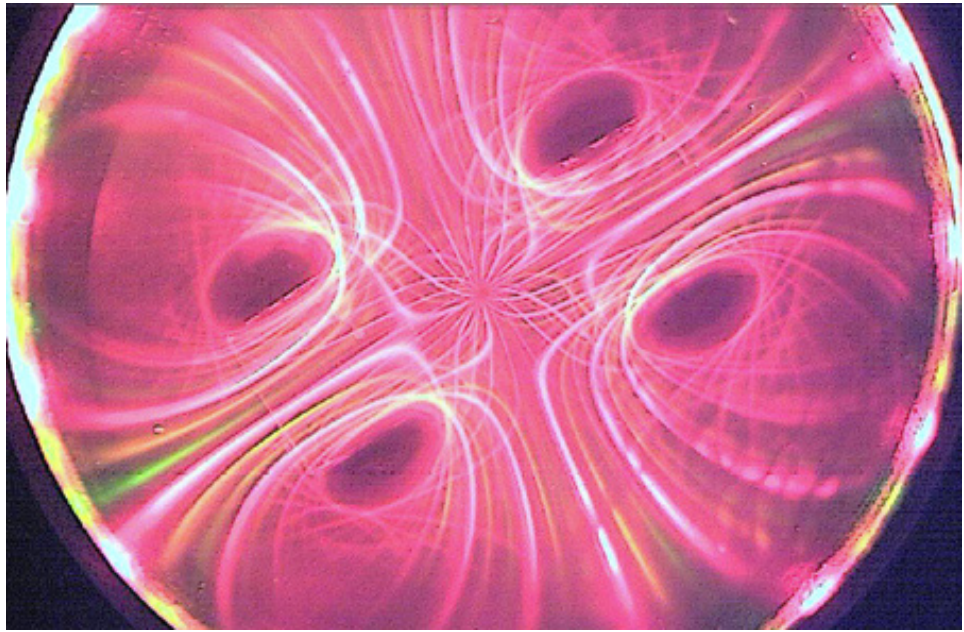

Figure 9: Photograph of two cube magnets. Left magnet has vertical north/south pole alignment and right magnet has vertical south/north pole alignment with different color LEDs around the perimeter of the lens.

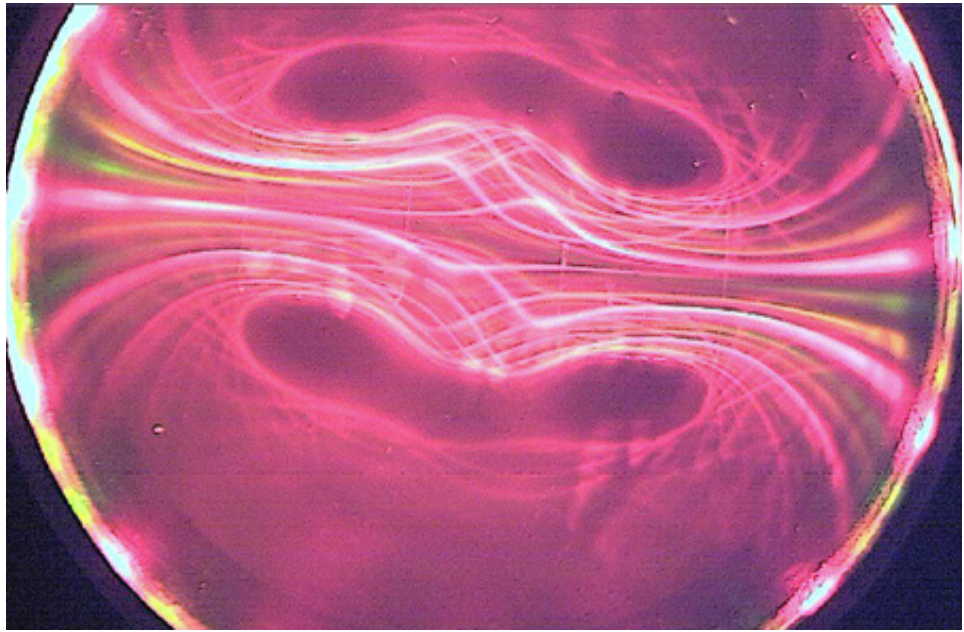

Figure 10: Photograph of two cube magnets. Profile view of both magnets with vertical north/south pole alignment with different color LEDs around the perimeter of the lens.

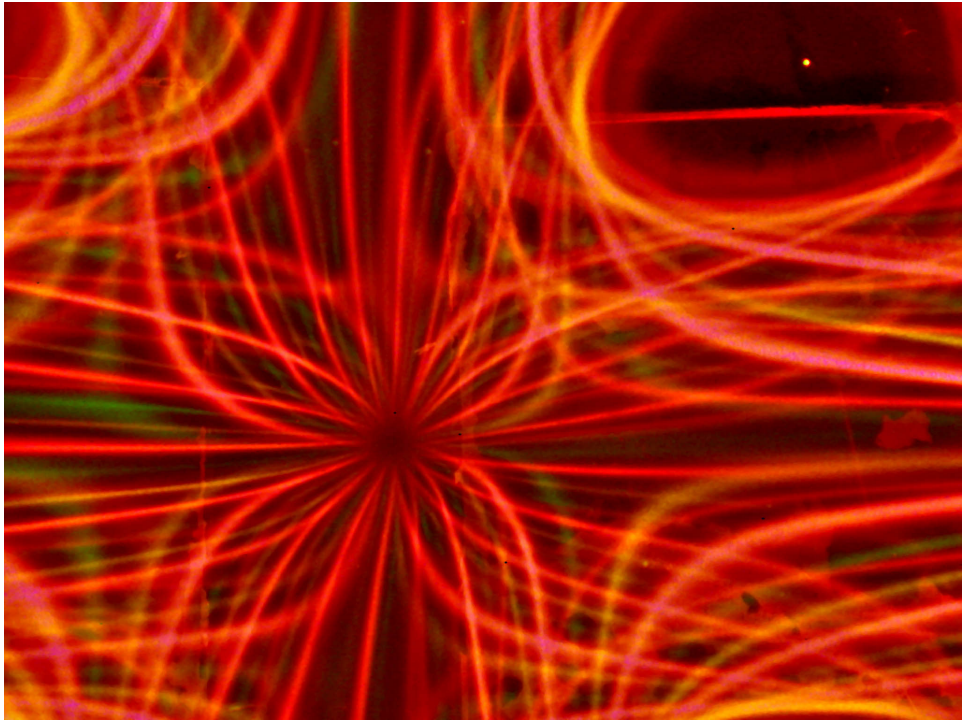

Figure 11: Zoomed photograph of two cube magnets. Left magnet has vertical north/south pole alignment and right magnet has vertical south/north pole alignment with different color LEDs around the perimeter of the lens.

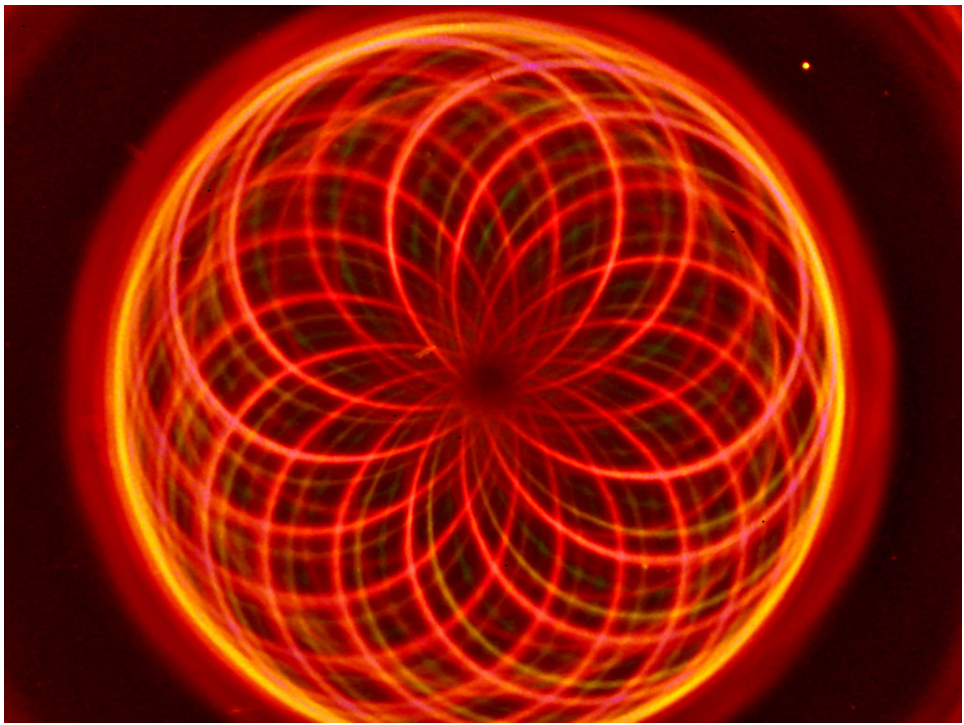

Figure 12: Photograph of 50.8mm ring magnet with the north pole facing the camera with different color LEDs around the perimeter of the lens.

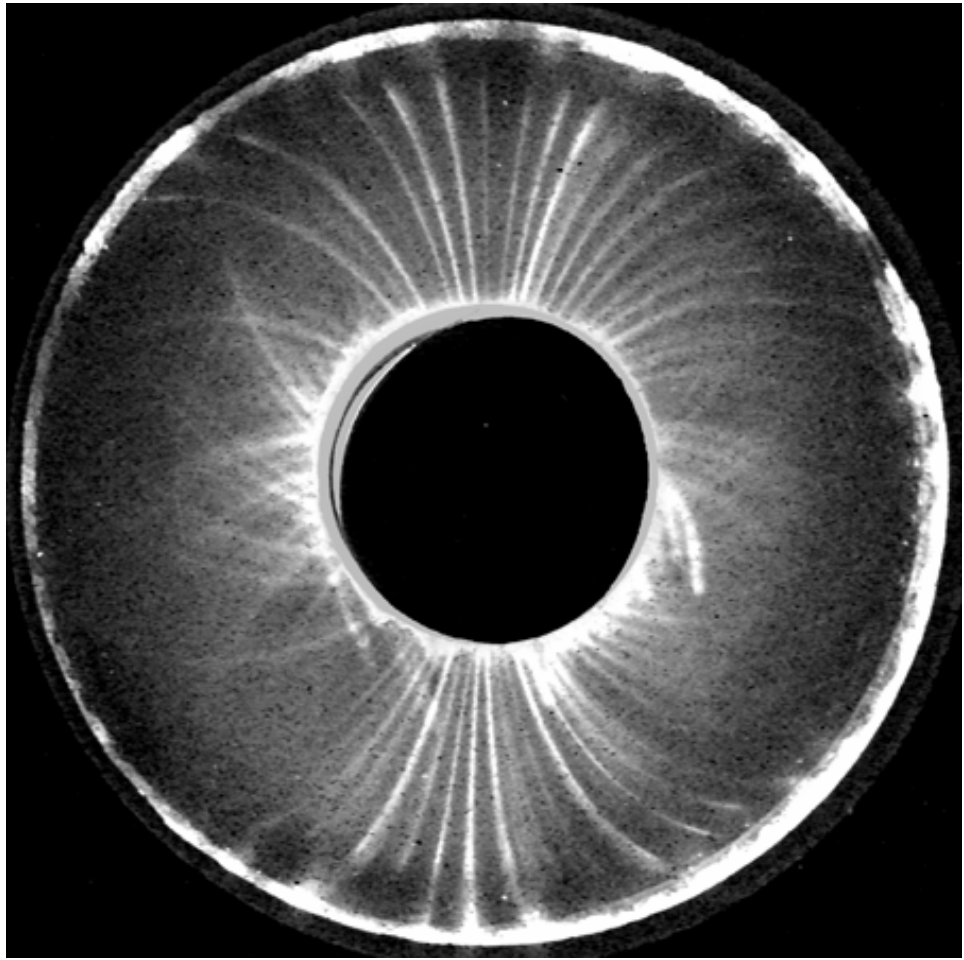

Figure S1: Source image #6

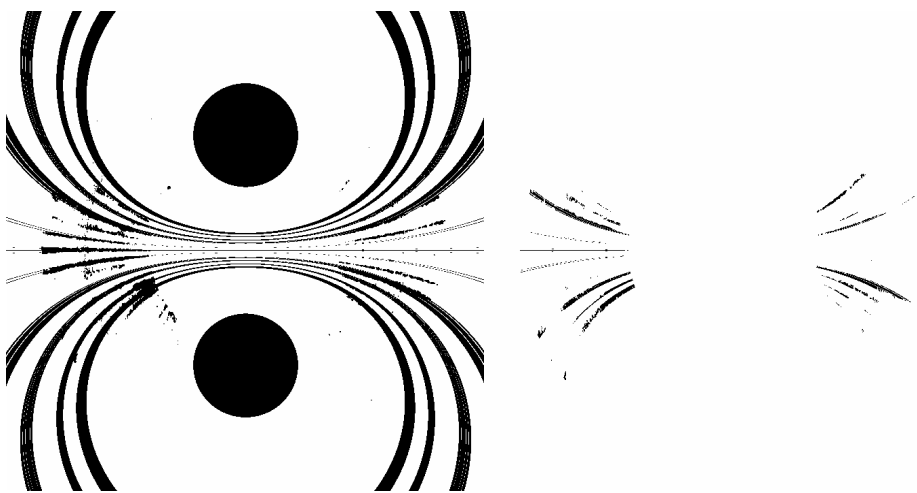

Figure F1: Fit of image #6 rotated 88 degrees, (experimental mask | idealized dipole) on Left, (experimental mask & idealized dipole) on Right

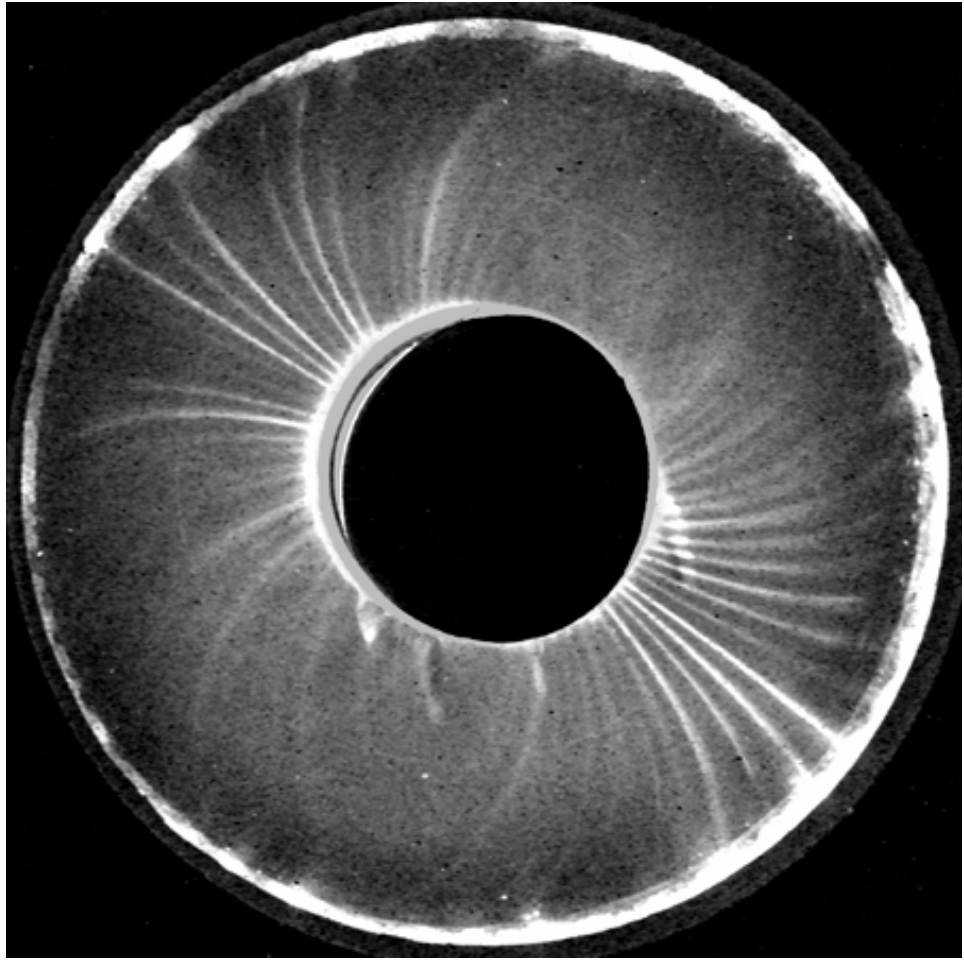

Figure S2: Source image #21

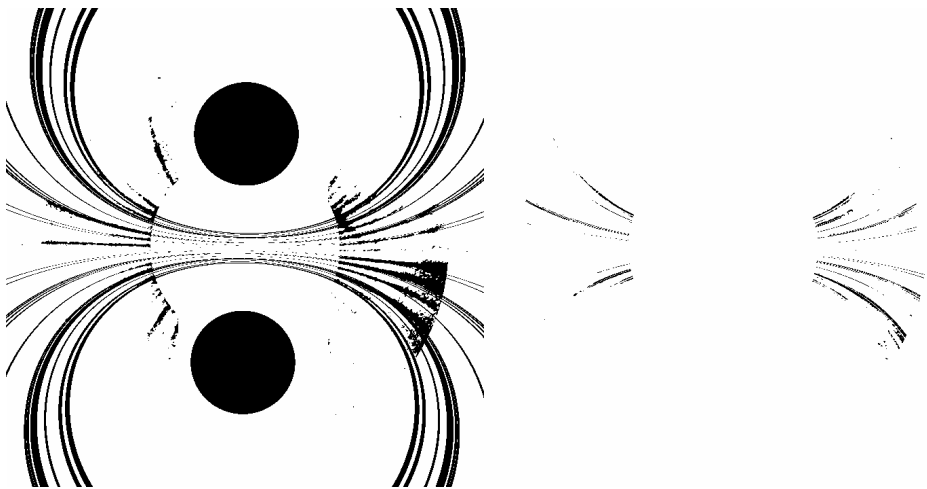

Figure F2: Fit of image #21 rotated 30 degrees, (experimental mask | idealized dipole) on Left, (experimental mask & idealized dipole) on Right

24